\documentclass[10pt, onecolumn, conference]{IEEEtran}

\usepackage{graphicx,url}
\usepackage[utf8]{inputenc}  
\usepackage{tabularx}

\usepackage{tikz}
\usetikzlibrary{positioning}
\title{A Taxonomy of Pix Fraud in Brazil: Attack Methodologies, AI-Driven Amplification, and Defensive Strategies}

\author{
\IEEEauthorblockN{
Glener Lanes Pizzolato,
Brenda Medeiros Lopes,
Claudio Schepke,
Diego Kreutz
}

\IEEEauthorblockA{
Graduate Program in Software Engineering (PPGES)\\
Advanced Computing Studies Laboratory (LEA) \& AI Horizon Labs\\
Federal University of Pampa (UNIPAMPA), Alegrete, Brazil\\
}
}

\begin{document} 

\maketitle

\begin{abstract}
This work presents a review of attack methodologies targeting Pix, the instant payment system launched by the Central Bank of Brazil in 2020. The study aims to identify and classify the main types of fraud affecting users and financial institutions, highlighting the evolution and increasing sophistication of these techniques. The methodology combines a structured literature review with exploratory interviews conducted with professionals from the banking sector. The results show that fraud schemes have evolved from purely social engineering approaches to hybrid strategies that integrate human manipulation with technical exploitation. The study concludes that security measures must advance at the same pace as the growing complexity of attack methodologies, with particular emphasis on adaptive defenses and continuous user awareness.
\end{abstract}

\begin{IEEEkeywords}
Pix, digital fraud, financial cybersecurity, social engineering attacks, instant payment systems, fraud taxonomy, behavioral analysis, attack methodologies, artificial intelligence in fraud, deepfakes, credential compromise, remote access malware, feature exploitation, phishing, scam detection, fraud mitigation, security controls, threat modeling.
\end{IEEEkeywords}

\section{Introduction}
Pix, launched by the Central Bank of Brazil (BCB) in November 2020, quickly became the most widely used means of payment in the country, surpassing traditional methods such as TED, DOC, payment slips, and card transactions.
Its large-scale adoption is due to its continuous availability, instant transfers, and operational simplicity through Pix keys \cite{bcb}.
Since its launch, the number of transactions and the amounts moved have grown rapidly, reaching tens of billions of operations per year \cite{bcb-estatisticas}.

The expansion of this system, however, has brought significant security challenges.  
Despite the mechanisms implemented by the BCB, incidents involving the leakage of registration data \cite{bcb-tabelaIncidentes}, social engineering scams, and express kidnappings have increased significantly after the introduction of Pix.  
In addition, the reduced use of physical cash has changed the criminal dynamics in the country, as pointed out by international studies that associate digital payments with lower rates of property crime.  
Recent cases, such as attacks involving the company Sinqia and the largest Pix-related theft ever recorded in Brazil, reinforce the critical nature of this issue.

In this context, the objective of this work is to identify and describe the main attempted attacks involving Pix transactions, to map their characteristics, and to propose a taxonomy that groups scams by type or similarity.  
It also seeks to analyze the role of Artificial Intelligence both in the execution of fraud and in the mitigation strategies adopted by financial institutions.  
To this end, attack methodologies, existing classifications, and security techniques implemented to protect transactions are investigated.

The central contribution includes:  
(a) a description of the methodologies employed in the main scams;  
(b) the mapping and classification of the attacks;  
(c) an analysis of the use of AI in offensive strategies;  
(d) a survey of the security techniques used by partner institutions; and  
(e) an investigation of the role of AI in defending against these attacks.  
These analyses provide a comprehensive view of the threat landscape surrounding Pix and support advances in the prevention, detection, and response to digital fraud in Brazil.

\section{Methodology}
The methodology of this work initially comprised a systematic survey of incidents and scams related to Pix, carried out through searches on news websites, specialized portals, and official leak statistics published by the Central Bank \cite{bcb-tabelaIncidentes}. 
After collection, the data underwent a cleaning process, eliminating duplicates and consolidating equivalent descriptions from different sources, which made it possible to create a uniform set of attacks for analysis.

Based on this consolidated set, a taxonomy was proposed, structured around three main pillars: motivation, medium, and execution. 
The classification of attacks was performed manually and validated with the assistance of three LLM models: GPT-\textit{4o} \cite{OpenAI_ChatGPT}, Gemini 2.5 Pro \cite{GoogleGemini}, and DeepSeek-V3 \cite{deepseekchat2024}. 
Each attack was categorized according to the motivation exploited, the medium used by the attacker, and the execution method employed.
Additionally, the potential use of artificial intelligence as a facilitating or amplifying element of the scam was evaluated, especially in stages involving the generation of convincing content or the creation of fake identities.

In parallel with the analysis of the scams, a survey was conducted on the security techniques adopted by financial institutions. 
This process involved open searches, consultations with LLMs, contact with representatives of banking agencies, conversations with security professionals, and interactions through institutional customer service channels. 
To ensure diversity and representativeness, thirteen institutions were selected, encompassing public, private, and digital banks, as well as cooperatives and highly capitalized institutions.
The techniques identified were also categorized according to whether or not they used artificial intelligence, allowing relationships to be established between defensive mechanisms and the previously classified attack methodologies.

\section{Results} \label{sec:mecanismos}

The survey identified 15 distinct methodologies of Pix-related scams, consolidated from 18 initial records after removal of duplicates.  
The attacks range from remote social engineering to structured fraud and direct physical interactions.  
Among the main ones are: tampered QR-code scam; scams involving physical interaction such as robberies and express kidnappings; fraud involving celebrities (the “Madonna scam”); fake receipt; fake scheduling; “ghost hand” scam (remote access); “Pix bug”; fake call centers; social engineering via WhatsApp; fake profiles; fake bank employee; WhatsApp cloning; wrong Pix; fake auction; and low-price scams.  
The analysis made it possible to organize them into a taxonomy based on the pillars of motivation, means, and execution, and later group them into four thematic categories summarized in Table~\ref{tab:grupos_golpes}.

\begin{enumerate}

\item \textbf{Tampered QR Code}: Criminals download legitimate broadcasts (such as NGO live streams) and rebroadcast them with a fraudulent QR Code to divert donations \cite{ataque-1}.

\item \textbf{Attack with Physical Interaction}: These include robberies, thefts, home invasions, and lightning kidnappings in which victims are coerced into making Pix transfers \cite{ataque-2-1, ataque-2-2}.

\item \textbf{Madonna Scheme}: Scammers send messages in the name of celebrities requesting donations via Pix, exploiting empathy and trust \cite{ataque-3}.

\item \textbf{Receipt Forgery (Pix)}: Forgery of Pix payment receipts with high realism, leading victims to believe in non-existent payments \cite{ataque-4}.

\item \textbf{False Scheduling Fraud}: The criminal sends a fake proof of a scheduled Pix transfer and requests the immediate return of the amount; after receiving it, they cancel the scheduling \cite{ataque-5}.

\item \textbf{Ghost Hand}: Also called ``remote access'', malware is installed on the cellphone that allows the scammer to operate the victim's account \cite{ataque-6}.

\item \textbf{The ``Pix Bug'' Scam}: Circulation of videos and messages falsely claiming a supposed error in the Pix system that would multiply transferred amounts \cite{ataque-7}.

\item \textbf{Fake Call Centers}: The victim is induced to call fraudulent numbers after receiving false alerts about alleged suspicious activities \cite{ataque-1}.

\item \textbf{Social Engineering Scam on WhatsApp}: The scammer uses the victim's photo and name on a new number and asks their contacts for money, claiming emergencies \cite{ataque-9}.

\item \textbf{Fake Profile Scam}: Creation of fake profiles on social networks using the victim's basic data to request Pix transfers from their contacts \cite{ataque-5}.

\item \textbf{Fake Bank Employee}: The criminal poses as a bank representative offering Pix support in order to gain access to the account \cite{ataque-11}.

\item \textbf{WhatsApp Cloning}: The scammer tricks the victim into providing the SMS activation code, clones the app, and asks the victim's contacts for Pix transfers \cite{ataque-12}.

\item \textbf{Wrong Pix Scam}: The victim receives an unexpected Pix transfer and is later induced to return the amount, without realizing the manipulation involved \cite{ataque-13}.

\item \textbf{Fake Auction Scam}: Fake auction websites require payment via Pix to secure non-existent products, exploiting urgency and unreal discounts \cite{ataque-14}.

\item \textbf{Low-Price Scheme}: After compromising social media accounts, criminals advertise products with attractive prices and request advance payment \cite{ataque-15}.

\end{enumerate}

\begin{table}[ht]
\centering
\caption{Thematic grouping of the mapped scams}
\label{tab:grupos_golpes}
\begin{tabular}{p{4cm}|p{10cm}}
\hline
\textbf{Group} & \textbf{Related Attacks} \\ \hline

Social Engineering Based on Authority and Trust
& [03] Madonna scam; 
[08] Fake call centers; 
[09] Social engineering via WhatsApp; 
[10] Fake profile; 
[11] Fake bank employee; 
[12] WhatsApp cloning \\ \hline

Social Engineering via Refund, Benefit, or Urgency
& [01] Tampered QR Code; 
[04] Fake receipt; 
[05] Fake scheduling; 
[07] Pix bug; 
[13] Wrong Pix transfer; 
[14] Fake auction; 
[15] Low prices \\ \hline

Attacks with Physical Interaction
& [02] Scams involving physical interaction: robbery, express kidnapping, extortion \\ \hline

Software- and Remote-Access-Based Attacks
& [06] Ghost hand / remote access \\ \hline

\end{tabular}
\end{table}

\noindent \textbf{Social Engineering Based on Authority and Trust.}
The attacks in this group exploit personal relationships, authority figures, or previously established trusted channels. 
Scams such as cloned WhatsApp accounts, fake profiles, fake bank representatives, and fake call centers operate on perceived credibility. 
These attacks tend to be highly effective, especially when combined with leaked data or personalized messages, creating an artificial trust scenario that increases the likelihood of human error.
\noindent \textbf{Social Engineering Through Refunds, Benefits, or Urgency.}
Scams in this group are structured around creating situations that demand quick actions from the victim.
Narratives involving refunds, financial opportunities, irresistible offers, or unexpected payments rely on emotional pressures such as greed, reciprocity, and urgency.
These are easily scalable attacks, often accompanied by forged receipts or fake pages, and they benefit from automation and personalization to reach large volumes of victims simultaneously.

\noindent \textbf{Attacks Involving Physical Interaction.}
Here, the fraud depends on direct coercion of the victim, as in robberies and express kidnappings.
Although execution is essentially in-person, there may be prior elements enabled by technology, such as target selection based on public data or inferred routines.
The risk is high, and countermeasures require integration between banking mechanisms for rapid blocking and public security policies.

\noindent \textbf{Software-Based Attacks and Remote Access.}
The “ghost hand” scam represents the most technical component, combining initial social engineering with malware-based execution.
It stands out for circumventing banking protections by obtaining direct control of the victim’s device, enabling transfers even without human interaction.
This type of attack tends to grow as automation and AI tools become more accessible.

Within this context, the taxonomy presented in this section organizes the fraud methodologies identified in the analyzed studies along three complementary dimensions. The first dimension, Motivation, captures the psychological triggers manipulated by fraudsters, including trust, fear, empathy, reciprocity, greed, and opportunity. These elements represent the emotional levers that enable social manipulation and shape the victim’s susceptibility to different types of persuasive strategies. The second dimension, Medium, specifies the channel through which the fraud is carried out, emphasizing remote social engineering, physical interaction, and software-based mechanisms. These vectors reflect both the diversity of operational environments and the increasing integration between social, physical, and digital components in contemporary fraud schemes. The third dimension, Execution, categorizes the concrete techniques used by malicious actors, such as deception, identity falsification, credential compromise, and the exploitation of legitimate system functionalities. This classification highlights how attackers transition from social influence to technical manipulation to complete the fraudulent process.

Taken together, these three dimensions make it possible to identify recurrent behavioral and operational patterns across different Pix fraud scenarios. They reveal that fraudulent schemes rarely rely on a single factor; instead, they emerge from the interplay between psychological influence, the channel through which the attacker interacts with the victim, and the technical or procedural method used to finalize the attack. By capturing this interplay, the taxonomy provides a structured framework for analyzing how attackers gradually progress from persuasion to exploitation, often beginning with emotional manipulation and advancing toward technical control or procedural abuse.

These dimensions also allow researchers to map how specific motivations align with particular attack vectors and execution strategies. For example, trust-based manipulation tends to favor remote social engineering as its primary medium, whereas fear-driven coercion is more frequently associated with physical interaction. Conversely, greed-oriented attacks exploit victims’ expectation of rapid gain, leveraging deceptive narratives that facilitate impersonation and credential compromise. Understanding these alignments is essential for identifying which vulnerabilities are psychological, which are technological, and which arise from gaps in financial procedures.

Furthermore, the taxonomy illustrates how modern Pix fraud schemes blend emotional manipulation, digital communication channels, and misuse of legitimate system functionalities to compromise victims in increasingly sophisticated ways. Instead of treating attacks as isolated events, the taxonomy contextualizes them within a broader ecosystem of fraud practices that evolve quickly, incorporate new digital tools, and exploit structural features of the payment system. This consolidated view supports the development of more effective detection methods, targeted awareness campaigns, and mitigation strategies aligned with real-world adversarial behaviors.

Table \ref{tab:taxonomia_consolidada} summarizes the most frequent combinations across the three dimensions, offering a high-level categorization of Pix fraud patterns based on the literature analyzed. These consolidated clusters represent the dominant configurations of motivation, medium, and execution found in real incidents and academic studies, providing a practical reference for research, analysis, and threat modeling.

\begin{table}[ht]
\centering
\renewcommand{\arraystretch}{1.2}
\resizebox{\textwidth}{!}{%
\begin{tabularx}{\textwidth}{p{3.2cm}|X|X}
\hline
\textbf{Motivation cluster} & \textbf{Medium} & \textbf{Execution pattern} \\ \hline
Trust and empathy based & Remote social engineering & Deception and impersonation \\ \hline
Fear and coercion based & Physical interaction (possibly combined with social engineering) & Credential compromise and impersonation \\ \hline
Greed and opportunity based & Remote social engineering & Deception and impersonation, sometimes combined with credential compromise \\ \hline
Mixed emotional triggers & Remote social engineering and software based mechanisms & Deception, impersonation and credential compromise \\ \hline
Trust and reciprocity based & Remote social engineering & Deception and functionality abuse (feature exploitation), often combined with impersonation \\ \hline
\end{tabularx}
}
\caption{Consolidated taxonomy of Pix fraud patterns}
\label{tab:taxonomia_consolidada}
\end{table}

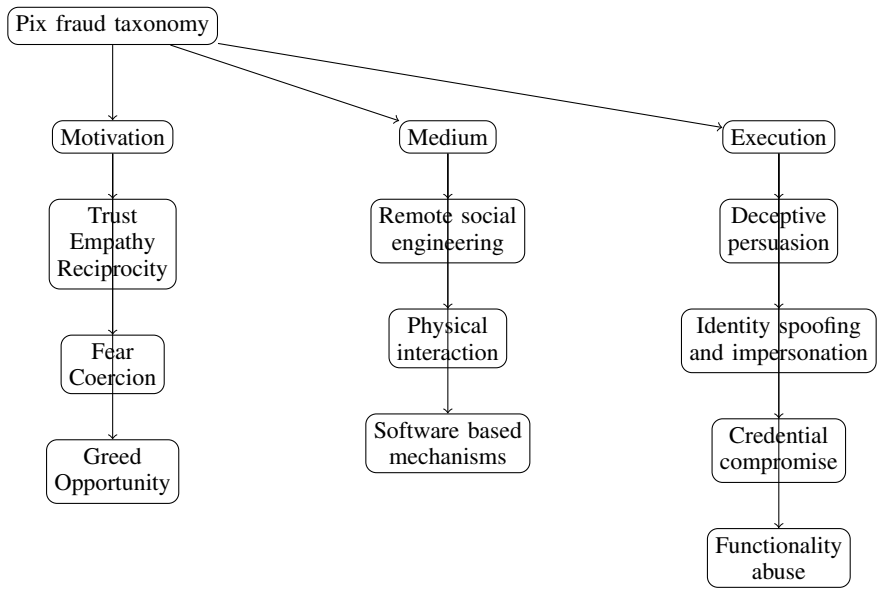
\begin{figure}[ht]
\centering
\begin{tikzpicture}[
  node distance=4mm and 12mm,
  every node/.style={rectangle, draw, rounded corners, align=center, font=\small, inner sep=3pt}
]

% Nó raiz
\node (root) {Pix fraud taxonomy};

% Nível 1: três dimensões
\node[below=10mm of root] (motivation) {Motivation};
\node[right=30mm of motivation] (medium) {Medium};
\node[right=30mm of medium] (execution) {Execution};

% Conexões raiz -> dimensões
\draw[->] (root) -- (motivation);
\draw[->] (root) -- (medium);
\draw[->] (root) -- (execution);

% Subnós de Motivation
\node[below=6mm of motivation] (mot1) {Trust \\ Empathy \\ Reciprocity};
\node[below=6mm of mot1]      (mot2) {Fear \\ Coercion};
\node[below=6mm of mot2]      (mot3) {Greed \\ Opportunity};

\draw[->] (motivation) -- (mot1);
\draw[->] (motivation) -- (mot2);
\draw[->] (motivation) -- (mot3);

% Subnós de Medium
\node[below=6mm of medium] (med1) {Remote social \\ engineering};
\node[below=6mm of med1]   (med2) {Physical \\ interaction};
\node[below=6mm of med2]   (med3) {Software based \\ mechanisms};

\draw[->] (medium) -- (med1);
\draw[->] (medium) -- (med2);
\draw[->] (medium) -- (med3);

% Subnós de Execution
\node[below=6mm of execution] (exe1) {Deceptive \\ persuasion};
\node[below=6mm of exe1]      (exe2) {Identity spoofing \\ and impersonation};
\node[below=6mm of exe2]      (exe3) {Credential \\ compromise};
\node[below=6mm of exe3]      (exe4) {Functionality \\ abuse};

\draw[->] (execution) -- (exe1);
\draw[->] (execution) -- (exe2);
\draw[->] (execution) -- (exe3);
\draw[->] (execution) -- (exe4);

\end{tikzpicture}
\caption{Conceptual taxonomy of Pix fraud}
\label{fig:pix_taxonomy}
\end{figure}

Figure~\ref{fig:pix_taxonomy} provides a visual synthesis of the proposed taxonomy by organizing the three core dimensions of Pix fraud into a structured hierarchical layout. The diagram arranges \textit{Motivation}, \textit{Medium}, and \textit{Execution} as distinct branches emerging from a central conceptual node, illustrating how psychological triggers, interaction channels, and operational techniques form interconnected components of a single fraudulent scheme. Each dimension is expanded into its corresponding subcategories, revealing recurring patterns such as trust-based manipulation, remote social engineering, and credential compromise. By presenting these elements within a unified visual framework, the figure clarifies how different stages and mechanisms combine to enable Pix fraud, underscoring that successful attacks typically arise from the coordinated interaction of emotional, technological, and procedural factors rather than isolated actions.

The investigation also confirmed that all mapped scams can be significantly amplified by modern artificial intelligence techniques. These include the generation of highly personalized persuasive messages, audio and video deepfakes capable of mimicking victims or trusted authorities, automated creation of convincing fake websites, construction of synthetic identities, and adaptive conversational bots that adjust their behavior in real time. Such capabilities reduce operational costs for criminals, increase the scalability of attacks, and weaken traditional detection mechanisms based on behavioral patterns or static signatures.

Interviews conducted with Banco do Brasil, Sicredi, and Banrisul further reinforce that social engineering frauds remain the predominant threat within the Pix ecosystem. Representatives from these institutions emphasize that their prevention strategies rely on a combination of multifactor authentication, continuous behavioral monitoring, artificial intelligence--driven detection models, and wide-reaching awareness and education campaigns targeting users. These measures collectively aim to reduce vulnerability to manipulation and increase early identification of suspicious activity.

The institutions consistently highlighted the speed and coordination of criminal operations as the primary barrier to recovering stolen funds. Fraudulent transfers are often executed and dispersed across multiple accounts within seconds, limiting the effectiveness of traditional intervention mechanisms. This scenario strengthens the argument for implementing faster and more automated blocking procedures, enhancing interbank cooperation, and investing in continuous user education to reduce susceptibility to initial social engineering vectors.

\section{Final Considerations and Future Work}

This study conducted a comprehensive review of attack methodologies involving the Pix system, mapping fifteen distinct scams and proposing a structured taxonomy that organizes these incidents according to motivation, medium, and execution.  
The analysis shows that most frauds occur through social engineering, highlighting the centrality of the human factor in the attack surface.  
The results indicate a clear evolution in criminal strategies, which have shifted from simple methods to increasingly sophisticated hybrid approaches, combining psychological manipulation, technical exploitation, and intensive use of automated mechanisms.  
The investigation also confirms that artificial intelligence currently plays a dual role: it significantly expands criminals’ offensive capabilities while simultaneously strengthening the defensive strategies adopted by financial institutions.

Interviews with Banco do Brasil, Sicredi and Banrisul reinforce that scams driven by social engineering remain the biggest operational challenge, since the speed of Pix transactions and the immediate dispersion of funds hinder recovery.
The institutions use multi-factor authentication, AI-based detection systems and awareness campaigns, but there is consensus that the speed of criminals and the emotional vulnerability of users still constitute critical bottlenecks.
In this way, this work contributes by systematizing a current and multifaceted overview of threats to Pix, providing a consistent conceptual basis for future studies and for the improvement of prevention, detection and response strategies.

As follow-up developments to this study, several avenues of investigation are proposed.
First, it is recommended to deepen the behavioral modeling of victims and attackers, enabling the development of predictive systems capable of identifying suspicious interactions before transactions are carried out.
In addition, future research may explore techniques for automatic detection of synthetic content (text, audio and video) generated by AI, considering the advancement of \textit{deepfakes} applied to financial scams.
Another promising direction involves studying mechanisms for contextual verification of transactions, combining real-time risk analysis and dynamic operating limits adjusted according to users’ historical patterns.

It is also suggested to investigate the impact of public policies and specific regulations that may reduce the attack surface in scenarios of social engineering and physical interaction, including unified emergency blocking protocols in the financial system.
Finally, the work can be expanded through the creation of a national \textit{dataset} of Pix-related scams, anonymized and standardized, serving as a basis for research in digital security, machine learning and the creation of fraud simulators for controlled testing.
These initiatives have the potential to significantly strengthen the resilience of the Pix ecosystem, keeping pace with the growing sophistication of the threats observed.

\bibliographystyle{ieeetr}
\bibliography{sbc-template}

\end{document}